\documentclass[twocolumn,aps,showpacs,prb]{revtex4-1}
\usepackage{graphicx}
\usepackage{epstopdf}
\usepackage{amsmath}
\usepackage{multirow}

\begin{document}

\title{The super-super exchange mechanism in iron-based antiperovskite chalco-halides}
\author{Kai Liu$^{1,2}$}\email{kliu@ruc.edu.cn}
\author{Zhong-Yi Lu$^{1,2}$}\email{zlu@ruc.edu.cn}

\affiliation{$^{1}$Department of Physics, Renmin University of China, Beijing 100872, China}
\affiliation{$^{2}$Beijing Key Laboratory of Opto-electronic Functional Materials $\&$ Micro-nano Devices, Renmin University of China, Beijing 100872, China}

\date{\today}

\begin{abstract}

By using the first-principles electronic structure calculations, we have systematically studied the magnetism in three recently synthesized iron-based antiperovskite chalco-halides: Ba$_3$(FeS$_4$)Cl, Ba$_3$(FeS$_4$)Br, and Ba$_3$(FeSe$_4$)Br. These compounds consist of edge-sharing Ba$Q_6$ ($Q$=Cl or Br) octahedra intercalated with isolated Fe$X_4$ ($X$=S or Se) tetrahedra. We find that even though the shortest distances between the nearest-neighboring Fe atoms in these three compounds already exceed 6 \AA, much larger than the bond length of a chemical bonding, they all remarkably show antiferromagnetic (AFM) coupling along $b$ axis with very weak spin-spin correlation along $a$ axis. Our study shows that the mechanism underlying this novel AFM coupling is such a new type of exchange interaction between the nearest-neighboring Fe-based super-moments mediated by Ba cations, which we call the super-super exchange interaction, in which each magnetic Fe atom partially polarizes its four nearest-neighboring $X$ atoms to form a super-moment through $p$-$d$ orbital hybridization and the $X$ atoms in neighboring Fe$X_4$ tetrahedra along $b$ axis antiferromagnetically couple with each others through the intermediate Ba cations. Different from the conventional superexchange, here it is cations rather than anions that mediate two neighboring super-moments. According to the calculated strength of the AFM coupling, we predict that among these compounds the highest AFM phase transition temperature $T_N$ may reach 110 K in Ba$_3$(FeSe$_4$)Br, in comparison with the observed $T_N$s of 84 K in Ba$_3$(FeS$_4$)Br and 95 K in Ba$_3$(FeS$_4$)Cl. 

\end{abstract}


\maketitle

\section{INTRODUCTION}

The mechanism of magnetic interactions is one of the most important topics in condensed matter physics. Previously, both direct exchange mechanism and indirect exchange mechanism, the latter including superexchange, double exchange, and Ruderman-Kittel-Kasuya-Yosida (RKKY) exchange, were successively proposed \cite{Stohr06}. The direct exchange is due to direct overlap of wavefuncitons, which is active in ferromagnetic metals. Among the indirect exchange mechanisms, the superexchange \cite{Kramers34, Anderson50} is adopted to explain the antiferromagnetic (AFM) interaction in transition metal oxides such as NiO; the double exchange \cite{Zener51, Anderson55, deGennes60} often applies to the ferromagnetic coupling in mixed-valence transition metal oxides such as perovskite La$_{1-x}$Sr$_x$MnO$_3$ ($0.16<x<0.5$) \cite{Tokura99}; with the RKKY exchange \cite{Ruderman54, Kasuya56, Yosida57}, the magnetic coupling in rare-earth’s intermetallic compounds, either ferromagnetic or antiferromagnetic depending on the distance from the local moment, can be well understood. Beyond fundamental scientific value, a thorough investigation of the magnetic interaction mechanism would be helpful for exploring novel magnetic materials and for accelerating their applications. 

A great number of perovskite oxides, with chemical formula of $AB$O$_3$ ($A$ = alkali metal, alkaline metal, rare earth etc; $B$ = transition metal; O = oxygen), have served as a playground for uncovering novel magnetic phenomena and for studying various magnetic mechanisms. To list some well-known magnetic properties in perovskites, the colossal magnetoresistance (CMR) in La$_{1-x}$Ca$_x$MnO$_3$ \cite{schiffer95, rao96, Tokura99book, dagotto01} and the multiferroic in BiFeO$_3$ \cite{wang03} were found in recent two decades. In addition, due to the interplay between lattice, charge, and spin, La$_{1-x}$Sr$_x$MnO$_3$ displays a complex magnetic phase diagram as functions of hole doping and temperature \cite{Tokura99}, demonstrating the sensitive modulation of its magnetic properties. At present, intensive studies on the abundant properties of perovskites are still in active progress \cite{Moure15}. 

The antiperovskites, which have the similar structures to perovskites but with the inverse cation and anion positions \cite{Krivovichev08}, are rare in nature. Nevertheless, some interesting magnetic properties such as magnetostriction \cite{Asano08} and piezomagnetic effects \cite{Lukashev08} have been discovered in metallic antiperovskites \cite{Tong12}. Recently, three new antiperovskite chaco-halides (Ba$_3$(FeS$_4$)Cl, Ba$_3$(FeS$_4$)Br, and Ba$_3$(FeSe$_4$)Br) have been synthesized \cite{Zhang15SR}. Although the shortest Fe-Fe distance is beyond 6.3 \AA, magnetic susceptibility measurements found unusual AFM phase transitions with temperatures $T_N$s about 95 K for Ba$_3$(FeS$_4$)Cl and 84 K for Ba$_3$(FeS$_4$)Br \cite{Zhang15SR}. Neutron scattering measurement and first-principles electronic structure calculations on one of these three compounds, Ba$_3$(FeS$_4$)Br, indicate that the AFM coupling induced by a new type of exchange along $b$ axis is responsible for the AFM phase transition \cite{Zhang15SR}. However, whether the same magnetic coupling applies to Ba$_3$(FeS$_4$)Cl and Ba$_3$(FeSe$_4$)Br, and what is the underlying effect influencing the strength of this AFM coupling between the largely distant neighboring Fe atoms, remain to be elucidated. A comprehensive understanding of these respects will provide us more insights into this novel exchange mechanism.

In this study, by using the first-principles electronic structure calculations, we have systematically investigated the electronic structures and magnetic properties of these three fresh antiperovskite chalco-halides: Ba$_3$(FeS$_4$)Cl, Ba$_3$(FeS$_4$)Br, and Ba$_3$(FeSe$_4$)Br. They show similar anisotropical magnetic couplings along different crystal axes. The novel mechanism of magnetic coupling between the largely distant neighboring Fe atoms along $b$ axis has been studied in detail. Counterintuitively, the calculations suggest that the strongest AFM coupling among these three compounds appears in Ba$_3$(FeSe$_4$)Br, which possesses the largest Fe-Fe distance. Further experimental measurement is required to verify our theoretical prediction.

\section{COMPUTATIONAL DETAILS}

The spin-polarized electronic structure calculations were performed with the projector augmented wave (PAW) method \cite{paw1, paw2} as implemented in the VASP package \cite{vasp1, vasp2, vasp3}. The exchange-correlation potentials were represented by the generalized gradient approximation (GGA) of Perdew-Burke-Ernzerhof (PBE) type \cite{pbe}. The kinetic energy cutoff of the plane-wave basis was set to be 350 eV. A 4$\times$6$\times$6 $k$-point mesh for the Brillouin zone sampling was used for the 1$\times$1$\times$1 orthogonal cell. The Gaussian smearing technique with a width of 0.05 eV was adopted for the Fermi level broadening. For structural optimization, both cell parameters and internal atomic positions were fully relaxed until all forces on atoms were smaller than 0.01 eV/\AA. The spin density and the charge difference density were analyzed at the equilibrium structure. 

\section{RESULTS AND ANALYSIS}

\begin{figure}[!t]
\includegraphics[angle=0,scale=0.4]{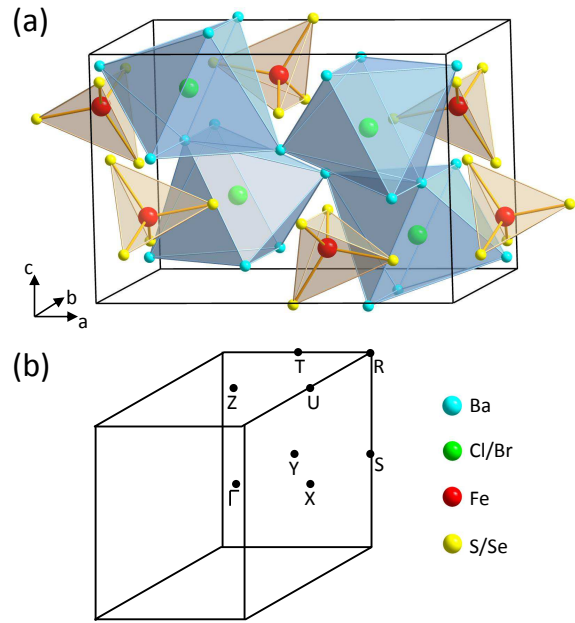}
\caption{(Color online) (a) Crystal structure of Ba$_3$(FeS$_4$)Cl, Ba$_3$(FeS$_4$)Br, and Ba$_3$(FeSe$_4$)Br. The cyan, green, red, and yellow balls denote the Ba, Cl/Br, Fe, and S/Se atoms, respectively. (b) Brillouin zone with high-symmetry $k$ points.}
\label{fig1}
\end{figure}

\begin{table}[!b]
\caption{Relative energies (in unit of eV/Fe) of different magnetic orders with respect to the nonmagnetic state for Ba$_3$(FeS$_4$)Cl, Ba$_3$(FeS$_4$)Br, and Ba$_3$(FeSe$_4$)Br.}
\begin{center}
\begin{tabular*}{8.5cm}{@{\extracolsep{\fill}} cccccccc}
\hline
\hline
$ $ & AFM1 & AFM2 & AFM3 & AFM4 & AFM5 & FM \\
\hline
Ba$_3$(FeS$_4$)Cl & -1.053 & -1.052 & -0.991 & -1.021 & -1.021 & -0.986 \\
Ba$_3$(FeS$_4$)Br & -1.077 & -1.078 & -1.019 & -1.047 & -1.047 & -1.015 \\
Ba$_3$(FeSe$_4$)Br & -1.135 & -1.134 & -1.057 & -1.094 & -1.094 & -1.050 \\
\hline \hline
\end{tabular*}
\end{center}
\end{table}

\begin{table*}[t]
\caption{Calculated lattice constants (in \AA), distances between the nearest-neighboring Fe atoms (in \AA) along $a$ axis ($d^{a}_{Fe-Fe}$) and $b$ axis ($d^{b}_{Fe-Fe}$), distances between the nearest $X$ ($X$=S or Se) atoms from the neighboring Fe$X_4$ tetrahedra (in \AA) along $a$ axis ($d^{a}_{X-X}$) and $b$ axis ($d^{b}_{X-X}$), local magnetic moments (in $\mu_B$), and band gaps $E_g$ (in eV) of Ba$_3$(FeS$_4$)Cl, Ba$_3$(FeS$_4$)Br, and Ba$_3$(FeSe$_4$)Br in their magnetic ground state (the AFM1 order). The lattice constants in the parentheses are experimental values from Ref. \onlinecite{Zhang15SR}.}
\begin{center}
\begin{tabular*}{16.5cm}{@{\extracolsep{\fill}} cccccccccccc}
\hline \hline
$ $ & $a$ & $b$ & $c$ & $d^{a}_{Fe-Fe}$ & $d^{b}_{Fe-Fe}$ & $d^{a}_{X-X}$ & $d^{b}_{X-X}$ & M$_{Fe}$ & M$_{S/Se}$ & $E_g$ \\
\hline
Ba$_3$(FeS$_4$)Cl & 12.44(12.25) & 9.58(9.54) & 8.52(8.42) & 6.32 & 6.31 & 4.11 & 3.81 & 3.40 & 0.15 & 0.865 \\
Ba$_3$(FeS$_4$)Br & 12.57(12.36) & 9.65(9.60) & 8.57(8.46) & 6.38 & 6.36 & 4.14 & 3.84 & 3.41 & 0.15 & 0.870 \\
Ba$_3$(FeSe$_4$)Br & 13.01(12.77) & 9.96(9.90) & 8.86(8.74) & 6.61 & 6.58 & 4.28 & 3.92 & 3.38 & 0.14 & 0.796 \\
\hline \hline
\end{tabular*}
\end{center}
\end{table*}

All three iron-based antiperovskite chaco-halides Ba$_3$(FeS$_4$)Cl, Ba$_3$(FeS$_4$)Br, and Ba$_3$(FeSe$_4$)Br take similar crystal structures (Figure 1). From Fig. 1(a), one can see that the Ba and $Q$ ($Q$=Cl or Br) atoms form corner-sharing Ba$Q_6$ octahedra while the Fe and $X$ ($X$=S or Se) atoms constitute isolated Fe$X_4$ tetrahedra intercalating between those octahedra. Along both $a$ and $b$ axes of the crystal, linking the nearest-neighboring Fe atoms makes up zigzag Fe chains, which can be identified more clearly when the other atomic species are omitted, as shown in Fig. 2. In Figure 2, the red and blue balls denote the spin-up and spin-down Fe atoms, respectively. There are four Fe atoms labeled Fe1, Fe2, Fe3, and Fe4 in the primitive cell [Fig. 2(a)], thus six possible magnetic orders with different spin orientations on the Fe atoms in primitive cell can be arranged (Fig. 2). For the AFM1, AFM2, and AFM3 orders, there are two spin-up and two spin-down Fe atoms in the primitive cell. As to the AFM4 and AFM5 orders, three spin-up and one spin-down Fe atoms show up. In the FM order, all Fe atoms adopt the same spin orientation. More complex spin patterns beyond the primitive cell are not considered.

\begin{figure}[!b]
\includegraphics[angle=0,scale=0.3]{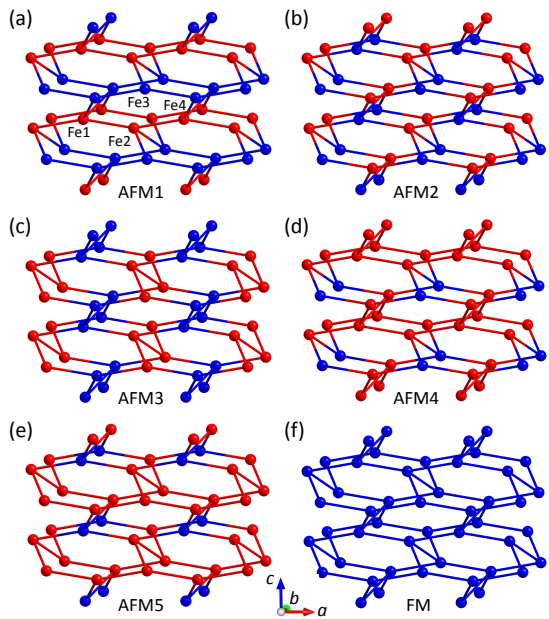}
\caption{(Color online) Six possible magnetic orders. There are four Fe atoms labeled Fe1, Fe2, Fe3, and Fe4 in the primitive cell. For clarity, other atomic species in the compound are omitted.}
\label{fig2}
\end{figure}

We have studied the energetics of these six possible magnetic orders for all three iron-based antiperovskite chaco-halides. After fully structural optimization, their relative energies with respect to the nonmagnetic order are listed in Table I. These compounds share the similar energy sequence among the different magnetic orders, while the AFM1 and AFM2 orders take the degenerate lowest energies. From the spin patterns shown in Fig. 2(a) and 2(b), one can discern that the AFM1 and AFM2 orders display the same AFM coupling between the nearest-neighboring Fe atoms along $b$ axis but demonstrate different spin orientations along $a$ axis. The spin pattern of the AFM1 order is in good accordance with the neutron powder diffraction measurement on  Ba$_3$(FeS$_4$)Br \cite{Zhang15SR}. Similarly, the energy difference between the AFM3 and FM orders are also very small. Once again, these two magnetic orders show the same Fe-Fe magnetic couplings along $b$ axis but distinct spin orientations along $a$ axis [Fig. 2(c) and 2(f)]. The AFM4 and AFM5 orders are also energetically degenerate. In fact, they are equivalent to each other by rotating 180 degrees around the $c$ axis. Overall, the ground states of all three iron-based antiperovskite chaco-halides are in the energetically degenerate AFM1 and AFM2 orders with the AFM coupling along $b$ axis, meanwhile the similar energies between the AFM1 and AFM2 orders as well as between the AFM3 and FM orders indicate that the magnetic coupling along $a$ axis is very weak.

\begin{figure}[!b]
\includegraphics[angle=0,scale=0.6]{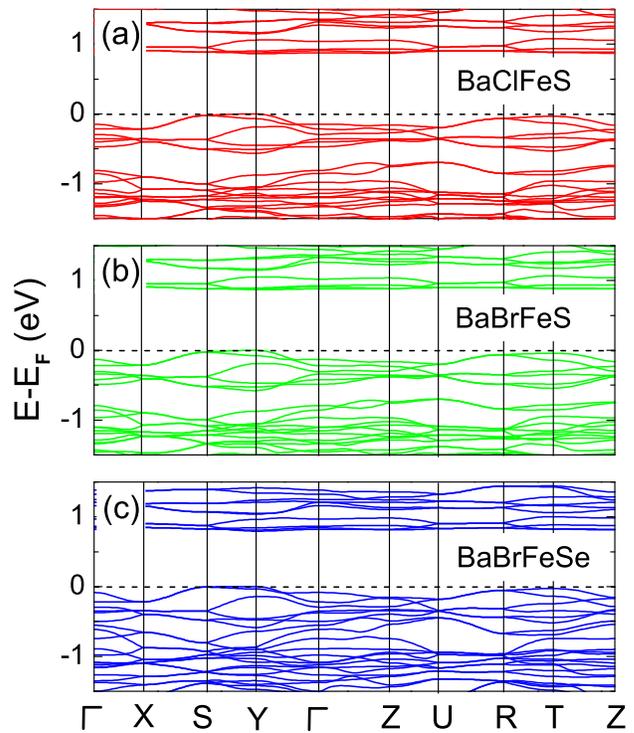}
\caption{(Color online) Band structures along high symmetry directions of Brillouin zone [Fig. 1(b)] for the AFM1 order of (a) Ba$_3$(FeS$_4$)Cl, (b) Ba$_3$(FeS$_4$)Br, and (c) Ba$_3$(FeSe$_4$)Br, respectively.}
\label{fig3}
\end{figure}

Table II lists the calculated lattice parameters, local magnetic moments, and energy gaps in the equilibrium structures of the magnetic ground states, i.e., the AFM1 orders, for Ba$_3$(FeS$_4$)Cl, Ba$_3$(FeS$_4$)Br, and Ba$_3$(FeSe$_4$)Br. The calculated lattice constants agree very well with the experimental values with errors no more than 2$\%$. For the internal atomic positions, the distances between the nearest-neighboring Fe atoms along all directions always exceed 6.3 \AA. Considering the semiconducting behavior of these antiperovskite chaco-halides, which show finite energy gaps (Fig. 3) and thus lack itinerant carriers, the existence of AFM coupling between the largely distant Fe atoms is very remarkable. By inspecting the calculated local moments, we get 3.38$\sim$3.41 $\mu_B$ on each Fe atom and 0.14$\sim$0.15 $\mu_B$ on each S/Se atom, respectively. The large moments on Fe atoms indicate that they are in the high-spin state, resulting from the Hund's rule coupling. 

The band gaps obtained from the band structures (Fig. 3) of the AFM1 orders are 0.865, 0.870, and 0.796 eV for Ba$_3$(FeS$_4$)Cl, Ba$_3$(FeS$_4$)Br, and Ba$_3$(FeSe$_4$)Br (Table II), respectively. Compared with the experimental values ($\sim$1.7 eV) \cite{Zhang15SR}, the underestimated band gaps from GGA calculations are due to the well-known derivative discontinuity of the exchange-correlation energy with respect to the number of electrons \cite{Perdew83}. Nevertheless, the quantitative difference between the calculated and measured band gaps does not influence our understanding on the magnetic coupling in these antiperovskite chaco-halides. Among them, the largest distance between the nearest-neighboring Fe atoms $\sim$6.6 \AA $ $ appears in Ba$_3$(FeSe$_4$)Br. Here we choose it as a prototypical system to investigate the spin interactions in the following.

For Ba$_3$(FeSe$_4$)Br in the AFM1 state, the spin distribution in real space can be visualized from its spin density as plotted in Figure 4. The magenta and cyan isosurfaces denote the respective spin-up and spin-down polarizations. Apparently, the Se atoms are polarized by their neighboring Fe atom through the $p$-$d$ hybridization between Se 4$p$ orbitals and Fe 3$d$ orbitals, yielding 0.14 $\mu_B$ on each Se (Table II).  Moreover, all atoms in the same FeSe$_4$ tetrahedron adopt the same spin orientation, behaving like a magnetic super-moment. In the AFM1 order, the spin densities of these FeSe$_4$ tetrahedra clearly show the ferromagnetic coupling along $a$ axis and the AFM coupling along $b$ axis. 

\begin{figure}[!t]
\includegraphics[angle=0,scale=0.33]{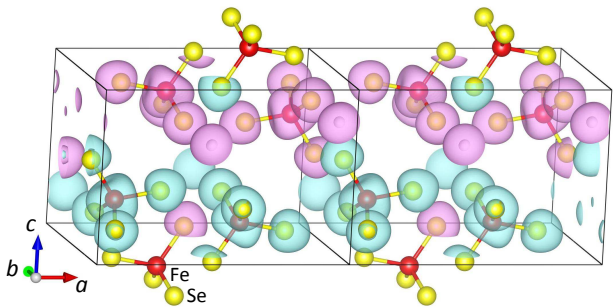}
\caption{(Color online) Spin density of Ba$_3$(FeSe$_4$)Br in the AFM1 order with the isosurface value of 0.02 e/\AA$^3$. The Ba and Br atoms are omitted for clarity.}
\label{fig4}
\end{figure}

Figure 5 displays the partial density of states (PDOS) of Ba$_3$(FeSe$_4$)Br in the AFM1 order. We choose a spin-up FeSe$_4$ tetrahedron for illustration. The strong $p$-$d$ hybridization between Fe 3$d$ orbitals and Se 4$p$ orbitals can be viewed from the common peaks in the same energy range as in Figure 5(a) and 5(b). For the Fe atom, five spin-up 3$d$ orbitals are fully occupied while five spin-down 3$d$ orbitals are occupied partially without obvious splitting. This results in a large local moment 3.38 $\mu_B$ on Fe atom. As to the Se atom, three spin-down 4$p$ orbitals are not fully occupied, rendering a small local moment on Se atom. In the above calculations, the degenerate total energies of the AFM1 and AFM2 orders (Table I) means that the magnetic coupling along $a$ axis is very weak. By checking the crystal structure [Fig. 1(a)], we can see that along $a$ axis a group of three nearest Se atoms between two neighboring FeSe$_4$ tetrahedra form a triangle, which can induce spin frustration easily. In addition, considering the larger Se-Se distance 4.28 \AA $ $ along $a$ axis than the one 3.92 \AA $ $ along $b$ axis (Table II), the weak coupling along $a$ axis can thus be understood. In contrast, along $b$ axis, a group of four nearest Se atoms in neighboring FeSe$_4$ tetrahedra constitute an approximate Se$_4$ rectangle bridged by two Ba atoms [Fig. 1(a)], which can be seen more clearly from Fig. 6(b). The PDOS from the 6$s$/6$p$ and 5$d$ orbitals of the Ba atom are plotted in Fig. 5(c) and 5(d), respectively. Compared with the PDOS of Fe and Se atoms, the PDOS of Ba atoms are much smaller, showing that the valence electrons of the Ba atom have been donated. 

\begin{figure}[!t]
\includegraphics[angle=0,scale=0.35]{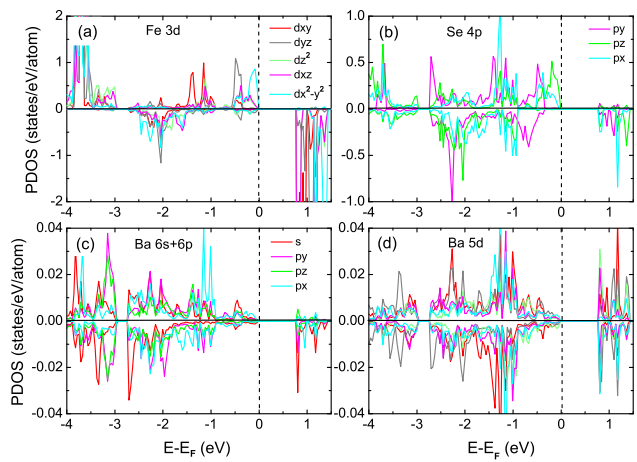}
\caption{(Color online) Partial density of states (PDOS) of Ba$_3$(FeSe$_4$)Br in the AFM1 state for (a) Fe 3$d$ orbitals, (b) Se 4$p$ orbitals, (c) Ba 6$s$ and 6$p$ orbitals, and (d) Ba 5$d$ orbitals. The legend for the five $d$ orbitals in panel (a) also applies to that in panel (d).}
\label{fig5}
\end{figure}

The interaction between the Ba and Se atoms along $b$ axis can be observed clearly from the charge difference density in Figure 6, in which one pair of FeSe$_4$ tetrahedra and two bridging Ba atoms are highlighted in Figs. 6(b) and 6(c). The distance between two Se atoms from neighboring FeSe$_4$ tetrahedra is 3.92 \AA, comparable with the intra-tetrahedron Se-Se distance 3.81 \AA. For one of the two Ba atoms locating above and below the center of the Se$_4$ rectangle [Fig. 6(b)], its distances from the Se atoms are around 3.41 and 3.72 \AA. From the charge difference density in Fig. 6(c), the electron depletion between the four Se atoms and the electron accumulation along the Ba-Se bonding direction can be discerned. This shows that the Se atoms in neighboring FeSe$_4$ tetrahedra are coupled with the bridging Ba atoms.

\begin{figure}[!t]
\includegraphics[angle=0,scale=0.33]{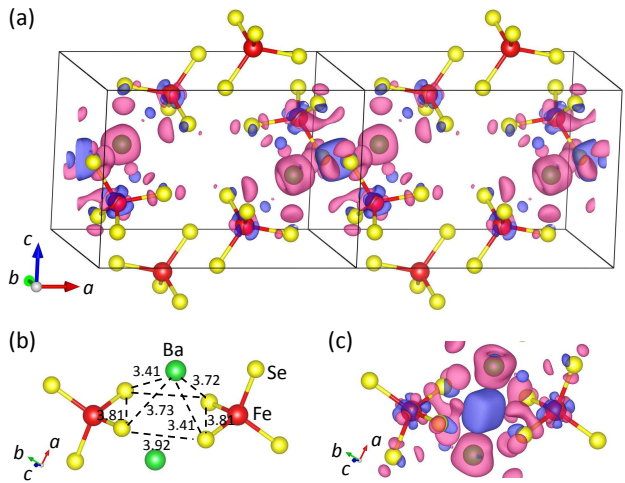}
\caption{(Color online) Charge difference density in isosurface of 0.02 e/\AA$^3$ for Ba$_3$(FeSe$_4$)Br in the AFM1 order. For clarity, one pair of neighboring FeSe$_4$ tetrahedra bridged by two Ba atoms along $b$ axis have been selected out in panels (b) and (c). The digits in panel (b) represent the distance in unit of \AA. The magenta and cyan isosurfaces denote the electron accumulation and depletion areas, respectively.}
\label{fig6}
\end{figure}

\section{DISCUSSION}

We now analyze why the Fe and Se atoms take the same spin orientations in the same FeSe$_4$ tetrahedron (Fig. 4). In the iron-based antiperovskite chalco-halides, the bonding of iron and chalcogen atoms falls in between the ionic and covalent characteristics, thus neither the Fe atom is in the exact +3 valence nor the Se atom is in the exact -2 valence. In fact, the respective majority spin channels of five Fe $3d$ orbitals and three Se $4p$ orbitals are fully occupied, while their minority spin channels are all partially filled [Figs. 5(a) and 5(b)]. For the Fe atom and one Se atom out of four in the same FeSe$_4$ tetrahedron, the $p$-$d$ hybridization between the Fe-$d_{yz}$ and Se-$p_y$, the Fe-$d_{xz}$ and Se-$p_x$, as well as the Fe-$d_{x^2-y^2}$ and Se-$p_z$ orbitals can be deduced from their common PDOS peaks just below the Fermi energy [Figs. 5(a) and 5(b)]. To lower the kinetic (band) energy due to electrons' hopping between Fe and Se orbitals in the Fe-Se bonding region, the electrons prefer taking the same spin polarization direction so that they become more extended.

The exchange between the largely distant Fe atoms found in our study \cite{Zhang15SR} is a new type of indirect exchange. In previous studies, there are three well-known forms of indirect exchange: superexchange, double exchange, and RKKY exchange \cite{Stohr06}. In case of the superexchange, the two nearest-neighboring magnetic moments are antiferromagnetically coupled via an intermediate nonmagnetic anion, whose two electrons in the same $p$-orbital bridging these two moments may virtually hop to the corresponding magnetic atoms respectively so that the kinetic energy may be further reduced, for example in compound La$_2$CuO$_4$.
In contrast, in case of the double exchange, the magnetic moments couple ferromagnetically with each other, due to the Hund's rule coupling between the itinerant $e_g$-orbital and localized $t_{2g}$-orbital electrons, for example in compound LaMnO$_3$. As to the RKKY exchange mechanism, the local moment of magnetic atom couples with the conduction electrons first, then the spin polarizations fade away from the localized moment with oscillating signs. Thus the spin information can spread over relatively long distance \cite{Stohr06}. In the Fe-based antiperovskite chaco-halides studied here, they are semiconductors without conduction electrons (Fig. 3). Moreover, the number of atoms between two magnetic Fe atoms are more than just one. Specifically, the magnetic Fe atom first polarizes its surrounding Se atoms in the same FeSe$_4$ tetrahedron via $p$-$d$ hybridization to form a super-moment, then the Se atoms couple antiferromagnetically with the corresponding Se atoms in a neighboring FeSe$_4$ tetrahedron along $b$ axis through such an exchange interaction mediated by the Ba cations, whose empty $s/p$-orbtials, bridging the Se atoms of neighboring FeSe$_4$ tetrahedra, will allow the polarized electrons hop in so as to further reduce the kinetic energy. We thus call this new type of exchange as super-super exchange. And we emphasize again that, unlike in conventional superexchange, here it is cations rather than an anion that mediate two neighboring super-moments. On the other hand, along the $a$ axis, the Se triangle formed by a group of three nearest Se atoms between two neighboring FeSe$_4$ tetrahedra [Fig. 1(a)] does not favor a long range magnetic order due to the spin frustration and larger Se-Se distance (Table II). To sum up, the local moments on Fe atoms are eventually antiferromagnetically coupled through the so-called super-super exchange mechanism along the $b$ axis.
 
The strength of super-super exchange along $b$ axis $J_b$ can be estimated from the energy differences between the AFM2 and AFM3 states. These energy differences render the $J_b$s of Ba$_3$(FeS$_4$)Cl, Ba$_3$(FeS$_4$)Br, and Ba$_3$(FeSe$_4$)Br as 31, 29, and 38 meV/Fe, respectively. The AFM phase transitions in those Fe-based antiperovskite chaco-halides closely correlate with this super-super exchange, as reflected in the observed $T_N$s of 95 K for Ba$_3$(FeS$_4$)Cl and 84 K for Ba$_3$(FeS$_4$)Br \cite{Zhang15SR}. If we assume the linear relationship between the AFM phase transition temperature and the super-super exchange, then a $T_N$ about 110 K would be expected for Ba$_3$(FeSe$_4$)Br, which is the highest among three compounds. However, the question arises if we further look at the crystal structural parameters in Table II. Along $b$ axis, the distances between the Fe atoms ($X$ atoms) of neighboring Fe$X_4$ tetrahedra are 6.31 (3.81), 6.36 (3.84), and 6.58 (3.92) \AA~in Ba$_3$(FeS$_4$)Cl, Ba$_3$(FeS$_4$)Br, and Ba$_3$(FeSe$_4$)Br, respectively. Apparently, the distances in Ba$_3$(FeSe$_4$)Br are the largest. Naively, one would think that Ba$_3$(FeSe$_4$)Br should possess the weakest super-super exchange. But this is controversial to the above calculated values of $J_b$. In order to understand this point, we compare the PDOS of Ba$_3$(FeSe$_4$)Br in Fig. 5 and that of Ba$_3$(FeS$_4$)Br in Fig. S8 in the supplementary materials of Ref. \onlinecite{Zhang15SR}. We find that for Fe atoms, their majority spin channels are both fully occupied and the minority spin channels partially occupied. The main difference between the Fe PDOS of these two compounds is that the peaks in the minority spin channel of Fe in Ba$_3$(FeSe$_4$)Br appears above -2.5 eV with respect to the Fermi level while the one in Ba$_3$(FeS$_4$)Br above -2.75 eV, indicating the spin-minority Fe $3d$ orbitals in Ba$_3$(FeSe$_4$)Br shift towards the Fermi level. Actually, this is influenced by the higher energies of Se 4$p$ orbitals than that of S 3$p$ orbitals, as can be discerned from the energies of occupied band tops in the minority spin channels for Se in Ba$_3$(FeSe$_4$)Br (-0.35 eV) [Fig. 5(b)] and for S in Ba$_3$(FeS$_4$)Br (-0.7 eV) [Fig. S8(b) in the supplementary materials of Ref. \onlinecite{Zhang15SR}]. The bigger atomic radius and modestly extended $p$ orbitals of Se than that of S induce stronger $p$-$d$ hybridization with the Fe $3d$ orbitals and stronger AFM superchange between neighboring Fe$X_4$ tetrahedra mediated by Ba atoms, and finally yield a greater super-super exchange strength in Ba$_3$(FeSe$_4$)Br.

\section{CONCLUSION}

By using the first-principles electronic structure calculations, we have investigated the spin-spin interactions in three recently synthesized iron-based antiperovskite chalco-halides: Ba$_3$(FeS$_4$)Cl, Ba$_3$(FeS$_4$)Br, and Ba$_3$(FeSe$_4$)Br, in which the shortest Fe-Fe distance is beyond 6.3 \AA. We find that the S/Se atoms are polarized by the magnetic Fe atom in the same FeS$_4$/FeSe$_4$ tetrahedron with the same spin orientation, resulting in a magnetic super-moment. The exchange interaction between such two nearest-neighboring super-moments mediated by the Ba cations along $b$ axis eventually results in antiferromagnetic coupling between the largely distant neighboring Fe spins, namely the super-super exchange interaction. This novel spin-spin coupling not only enriches our knowledge on the types of magnetic interactions in condensed matters, but also provides a possible approach to transfer spin information for a long distance with less magnetic atoms.

\begin{acknowledgments}

We wish to thank Fu-Qiang Huang and Xian Zhang for helpful discussions. This work was supported by the National Natural Science Foundation of China (Grants 11190024 and 91421304), the Fundamental Research Funds for the Central Universities, and the Research Funds of Renmin University of China. Computational resources have been provided by the Physical Laboratory of High Performance Computing at RUC.

\end{acknowledgments}


\begin{thebibliography}{}

\bibitem{Stohr06} J. St\"ohr and H. C. Siegmann, Magnetism: From fundamentals to nanoscale dynamics (Springer, Berlin, 2006).

\bibitem{Kramers34} H. A. Kramers, Physica {\bf 1}, 182 (1934). 

\bibitem{Anderson50} P. W. Anderson, Phys. Rev. {\bf 79}, 350 (1950).

\bibitem{Zener51} C. Zener, Phys. Rev. {\bf 82}, 403 (1951).

\bibitem{Anderson55} P. W. Anderson and H. Hasegawa, Phys. Rev. {\bf 100}, 675 (1955).

\bibitem{deGennes60} P. G. de Gennes, Phys. Rev. {\bf 118}, 141 (1960).

\bibitem{Tokura99} Y. Tokura and Y. Tomioka, J. Magn. Magn. Mater. {\bf 200}, 1 (1999).

\bibitem{Ruderman54} M. A. Ruderman and C. Kittel, Phys. Rev. {\bf 96}, 99 (1954).

\bibitem{Kasuya56} T. Kasuya, Prog. Theor. Phys. {\bf 16}, 45 (1956).

\bibitem{Yosida57} K. Yosida, Phys. Rev. {\bf 106}, 893 (1957).

\bibitem{schiffer95} P. Schiffer, A. P. Ramirez, W. Bao, and S.-W. Cheong, Phys. Rev. Lett. {\bf 75}, 3336 (1995).

\bibitem{rao96} C. N. R. Rao, A. K. Cheetham, and R. Mahesh, Chem. Mater. {\bf 8}, 2421 (1996).

\bibitem{Tokura99book} Y. Tokura, Colossal Magnetoresistive Oxides (Gordon and Breach, London, 1999).

\bibitem{dagotto01} E. Dagotto, T. Hotta, and A. Moreo, Phys. Rep. {\bf 344}, 1 (2001).

\bibitem{wang03} J. Wang, J. B. Neaton, H. Zheng, V. Nagarajan, S. B. Ogale, B. Liu, D. Viehland, V. Vaithyanathan, D. G. Schlom, U. V. Waghmare, N. A. Spaldin, K. M. Rabe, M. Wuttig, and R. Ramesh, Science  {\bf 299}, 1719 (2003). 

\bibitem{Moure15} C. Moure and O. Pena, Prog. Solid State Chem. {\bf 43}, 123 (2015).

\bibitem{Krivovichev08} S. V. Krivovichev, Zeitschrift f\"ur Kristallographie {\bf 223}, 109 (2008). 

\bibitem{Asano08} K. Asano, K. Koyama, and K. Takenaka, Appl. Phys. Lett. {\bf 92}, 161909 (2008).

\bibitem{Lukashev08} P. Lukashev, R. F. Sabirianov, and K. Belashchenko, Phys. Rev. B {\bf 78}, 184414 (2008).

\bibitem{Tong12} P. Tong and Y. P. Sun, Adv. Conden. Matter Phys. doi:10.1155/2012/903239 (2012).


\bibitem{Zhang15SR} X. Zhang, K. Liu, J.-Q. He, H. Wu, Q.-Z. Huang, J.-H. Lin, Z.-Y. Lu, and F.-Q. Huang, Sci. Rep. {\bf 5}, 15910 (2015).

\bibitem{paw1} P. E. Bl\"ochl, Phys. Rev. B {\bf 50}, 17953 (1994).

\bibitem{paw2} G. Kresse and D. Joubert, Phys. Rev. B. {\bf 59}, 1758 (1999).

\bibitem{vasp1} G. Kresse and J. Hafner, Phys. Rev. B {\bf 47}, 558 (1993). 

\bibitem{vasp2} G. Kresse and J. Furthm\"uller, Comp. Mater. Sci. {\bf 6}, 15 (1996). 

\bibitem{vasp3} G. Kresse and J. Furthm\"uller, Phys. Rev. B {\bf 54}, 11169 (1996).

\bibitem{pbe} J. P. Perdew, K. Burke, and M. Ernzerhof, Phys. Rev. Lett. {\bf 77}, 3865 (1996).

\bibitem{Perdew83} J. P. Perdew and M. Levy, Phys. Rev. Lett. {\bf 51}, 1884 (1983).

\end{thebibliography}
\end{document}